%% file: main.tex
\documentclass[a4paper, 10pt, twocolumn]{article} 
\input{gamma26-template}
\usepackage{amssymb}
\usepackage{enumitem}

\title{
    Unveiling the Extreme Universe with Recent IACT Discoveries
}

\author{
	\authorstyle{Rub{\'e}n L{\'o}pez-Coto\textcolor{GammaDarkGreen}{\textsuperscript{1}}}
	\newline\newline
    {\footnotesize
        \textsuperscript{\color{GammaDarkGreen}1}\institution{Instituto de Astrof{\'i}sica de Andaluc{\'i}a---CSIC, Glorieta de la Astronom{\'i}a s/n, E-18008 Granada, Spain}
    }
}

\date{\today}
\newcommand{\website}{\href{https://plan.events.mpg.de/event/543/}{Proceedings of the 9th Heidelberg Gamma-Ray Symposium}} 

\begin{document}
\linespread{1.00}\selectfont
\setlength{\parskip}{0.14ex}
\sansmath\maketitle
\thispagestyle{fancy}

\lettrineabstract{Very-high-energy (VHE, $E \gtrsim 50$\,GeV) $\gamma$-ray astronomy has evolved into an indispensable pillar of modern astroparticle physics, offering a direct view of non-thermal particle acceleration, cosmic-ray origins, and fundamental physics. Ground-based Imaging Atmospheric Cherenkov Telescopes (IACTs) have driven this revolution over the past two decades. In this plenary review, presented at the 9th Heidelberg International Symposium on High-Energy Gamma-Ray Astronomy (Gamma 2026), I provide a synthesis of recent breakthrough physics discoveries achieved by current IACT arrays (H.E.S.S., MAGIC, and VERITAS) alongside early scientific achievements from the Large-Sized Telescope prototype (LST-1) of the Cherenkov Telescope Array Observatory (CTAO). Key highlights include the definitive establishment of recurrent novae as transient cosmic-ray proton accelerators via RS~Ophiuchi; spatially resolved multi-TeV emission from the relativistic jet termination shocks of microquasars; the discovery of a $20$\,TeV pulsed radiation component from the Vela pulsar, challenging established magnetospheric emission paradigms; precision measurements of the Galactic Center diffuse ridge spectrum favoring a high-energy cutoff; the detection of distant flat-spectrum radio quasars at $z \sim 1$ (OP~313 and PKS~0346-27), providing tight cosmological constraints on the Extragalactic Background Light; multi-wavelength afterglow detections of $\gamma$-ray bursts including GRB~221009A; and world-leading joint indirect dark matter limits from dwarf spheroidal galaxy and Galactic Center halo analyses. Together with algorithmic advances in deep learning reconstruction and the ongoing construction of CTAO, these milestones demonstrate the enduring vitality and future trajectory of ground-based $\gamma$-ray astrophysics.}

\section{Introduction: The Ground-Based VHE Landscape}
\label{sec:intro}

The non-thermal Universe is populated by extreme astrophysical environments where charged particles are accelerated to ultra-relativistic energies. When interacting with ambient gas, magnetic, or radiation fields, these relativistic leptons and hadrons produce high-energy $\gamma$-ray photons via synchrotron radiation, relativistic bremsstrahlung, inverse Compton (IC) scattering, and the inelastic production and subsequent decay of neutral pions. Because $\gamma$ rays carry no electric charge, they propagate along undeflected trajectories through interstellar and intergalactic magnetic fields, providing an unperturbed directional pointer directly back to their cosmic acceleration engines. As such, VHE $\gamma$ rays serve as direct electromagnetic counterparts to the cosmic-ray flux and provide critical complementary insights to astrophysical neutrino observations by facilities such as IceCube and KM3NeT.

While space-borne detectors such as the \textit{Fermi} Large Area Telescope (\textit{Fermi}-LAT) have surveyed the $\gamma$-ray sky below $\sim 100$\,GeV with wide fields of view and continuous all-sky coverage, the steeply falling power-law spectra typical of non-thermal cosmic sources ($\mathrm{d}N/\mathrm{d}E \propto E^{-\Gamma}$ with $\Gamma \gtrsim 2$) impose a statistical barrier on satellite collection areas ($\lesssim 1\text{ m}^2$) at higher energies. Ground-based Imaging Atmospheric Cherenkov Telescopes (IACTs) overcome this limitation by utilizing the Earth's atmosphere as a calorimeter~\cite{Weekes1989,Hillas1985}. An incident VHE $\gamma$~ray initiates an extensive electromagnetic air shower in the upper atmosphere, whose ultra-relativistic electron and positron pairs generate a collimated cone of optical Cherenkov light. Segmented optical reflectors focus this brief ($\sim$ few\,ns) optical flash onto fast, high-pixelized cameras, achieving instantaneous effective collection areas of $A_{\rm eff} \sim 10^5\text{ m}^2$ on the ground.

Over the past two decades, the current generation of stereoscopic IACT facilities -the High Energy Stereoscopic System (H.E.S.S.) in Namibia, the Major Atmospheric Gamma Imaging Cherenkov (MAGIC) telescopes on La Palma, and the Very Energetic Radiation Imaging Telescope Array System (VERITAS) in Arizona- has transformed VHE $\gamma$-ray astronomy from an exploratory pursuit into a precision observational discipline. Operating across four orders of magnitude in energy ($\sim 20$\,GeV to $>100$\,TeV), these observatories have revealed over 300 distinct VHE emitters spanning diverse Galactic and extragalactic source classes~\cite{HESSGPS2005}. Stereoscopic reconstruction provides excellent angular resolution ($\lesssim 0.05^\circ$--$0.1^\circ$), energy resolution of $10\%$--$15\%$, and rejection of $>99\%$ of the hadronic cosmic-ray background. These pointed instruments provide an essential complement to wide-field ground-based particle arrays such as HAWC, LHAASO, and the future Southern Wide-field Gamma-ray Observatory (SWGO), offering deep spectroscopic and morphological scrutiny of selected targets.

Today, the discipline stands at the threshold of a generational transition to the Cherenkov Telescope Array Observatory (CTAO), which is deploying two world-class arrays: CTAO-North at the Roque de los Muchachos Observatory on La Palma, Spain, and CTAO-South near Paranal in Chile~\cite{CTAScience2019}. CTAO-North's Large-Sized Telescope prototype (LST-1, $23$-m diameter) is already operating routinely on La Palma, delivering high sensitivity down to $\approx 20$\,GeV and rapid repositioning capabilities ($<20$\,s) that are opening unprecedented observational windows into transient and distant cosmic phenomena~\cite{LST1Perf2023}. 

In this proceeding, I review the most impactful scientific results recently achieved by current IACT arrays and LST-1. The paper is structured as follows: in Sect.~\ref{sec:novae}, I discuss the establishment of recurrent novae as transient hadronic accelerators; in Sect.~\ref{sec:microquasars}, I review particle acceleration and cooling in microquasar jets; in Sect.~\ref{sec:pulsars}, I present the 20\,TeV challenge in pulsar magnetospheres; in Sect.~\ref{sec:galactic_center}, I evaluate the PeVatron status of the Galactic Center; in Sect.~\ref{sec:extragalactic}, I discuss distant blazars and EBL constraints; in Sect.~\ref{sec:grbs}, I cover recent GRB afterglow observations; in Sect.~\ref{sec:dark_matter}, I review indirect dark matter searches; and in Sect.~\ref{sec:outlook}, I highlight the outlook for CTAO and deep learning methods, before concluding in Sect.~\ref{sec:conclusions}.

\section{Thermonuclear Novae as Transient Hadronic Accelerators}
\label{sec:novae}

Classical and recurrent novae are catastrophic stellar eruptions occurring in close binary systems where a white dwarf (WD) accretes hydrogen-rich material from a non-degenerate companion star until a thermonuclear runaway is triggered in the degenerate accreted envelope. Historically regarded as purely thermal optical and ultraviolet transients, novae were recognized as particle accelerators following \textit{Fermi}-LAT detections of GeV $\gamma$ rays produced in internal shocks. However, whether novae could accelerate particles beyond tens of GeV into the VHE regime remained an open question until the 2021 outburst of the symbiotic recurrent nova RS~Ophiuchi (RS~Oph).

\begin{figure}[t]
    \centering
    \includegraphics[width=\columnwidth]{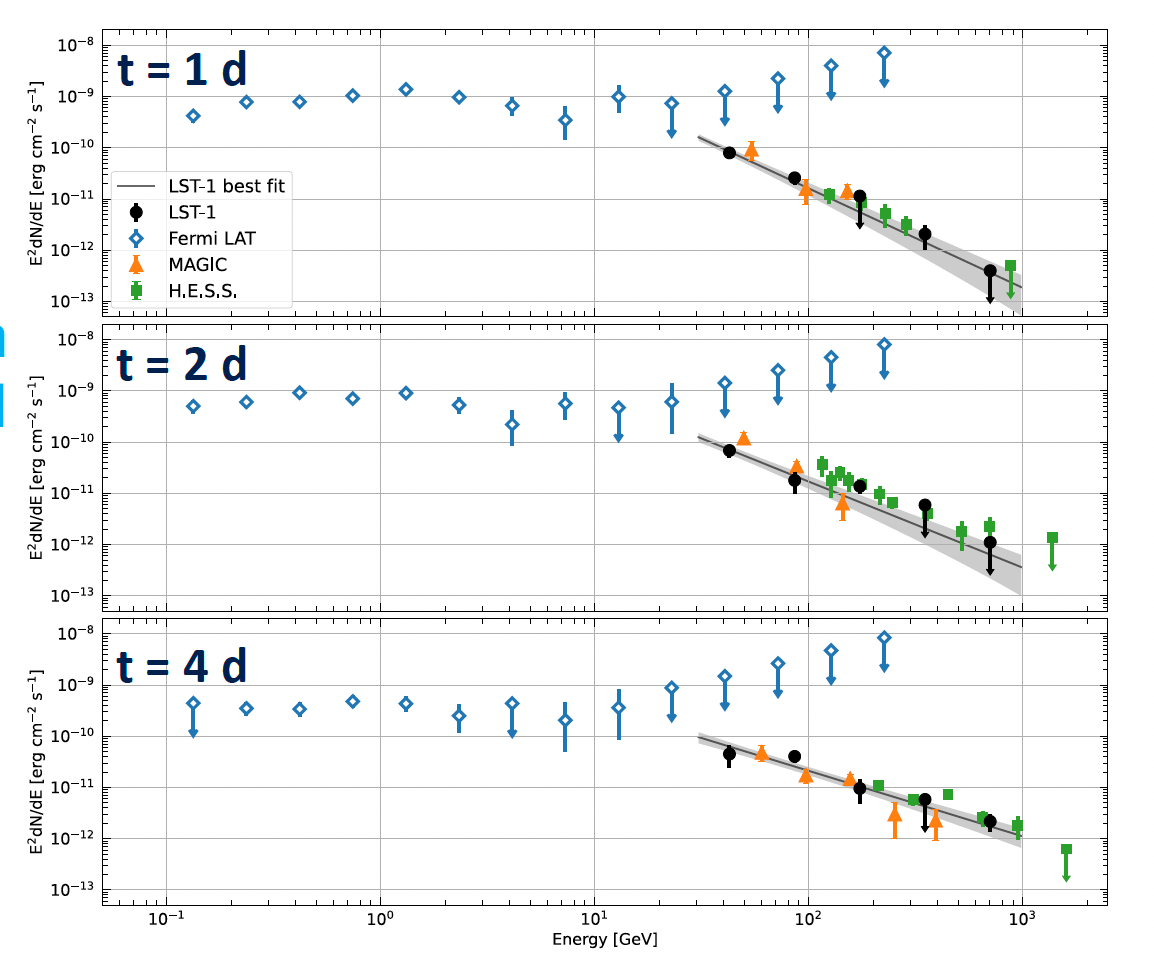}
    \caption{Differential energy spectra of the recurrent nova RS~Ophiuchi during its 2021 outburst, measured with LST-1 on days 1, 2, and 4 post-outburst together with contemporaneous \textit{Fermi}-LAT measurements. The consistent spectral shape across consecutive nights reveals sustained proton acceleration in the expanding blast wave with a smooth high-energy turnover. Original image from: \cite{LSTRSOph2025}.}
    \label{fig:rs_oph_spectra}
\end{figure}

RS~Oph consists of a massive white dwarf ($\approx 1.2$--$1.4\,M_\odot$) orbiting an M-type red giant donor in a $454$-day binary orbit, experiencing recurrent eruptions every $15$--$20$\,years. Following its eruption on 2021 August 8, ground-based IACTs triggered rapid Target of Opportunity observations. RS~Oph was detected at VHE energies by H.E.S.S.~\cite{HESSRSOph2022}, MAGIC~\cite{MAGICRSOph2022}, and LST-1~\cite{LSTRSOph2025}, definitively establishing novae as a novel class of VHE $\gamma$-ray emitters and extending their known spectral reach by three orders of magnitude.

As displayed in Fig.~\ref{fig:rs_oph_spectra}, LST-1 performed dedicated observations of RS~Oph across consecutive nights (days 1, 2, and 4 post-outburst), measuring differential spectra extending from $\sim 50$\,GeV to several hundred GeV~\cite{LSTRSOph2025}. The measured spectral index and normalization remained remarkably stable during these initial days, forming a multi-day plateau before transitioning into a steady temporal decay that tracked the multi-wavelength decline seen in GeV and optical observations.

The multi-observatory data provide clear, decisive physical insights into the nature of particle acceleration in novae:
\begin{itemize}[leftmargin=*,itemsep=0ex,parsep=0ex,topsep=0.2ex]
    \item The broadband $\gamma$-ray emission is naturally explained by a hadronic scenario, where protons are accelerated via diffusive shock acceleration (DSA) in the blast wave expanding into the dense circumstellar wind of the red giant companion. The parent proton spectrum follows a power law with a smooth exponential cutoff at several hundred GeV. 
    
    In contrast, purely leptonic models based on inverse Compton scattering fail to account for the multi-day plateau. In a leptonic scenario, the intense optical radiation field of the expanding nova photosphere serves as the primary seed photon target for IC scattering. As the photosphere expands, its radiation density drops steeply, which would produce an immediate, precipitous decrease in VHE flux that is ruled out by the observations. Conversely, the hadronic interaction timescale with the dense red giant wind gas naturally matches the dynamical shock expansion timescale, explaining the concurrent decay observed across optical, GeV, and VHE bands.
    \item The total energy channeled into relativistic protons is estimated at $W_p \approx (1$--$3) \times 10^{43}$\,erg, corresponding to $\approx 0.1\%$--$1\%$ of the total nova kinetic energy budget ($\approx 10^{45}$\,erg)~\cite{HESSRSOph2022,MAGICRSOph2022}. Integrating over the Galactic nova rate ($\sim 30$--$50\text{ yr}^{-1}$) indicates that novae contribute $<0.2\%$ of the Galactic cosmic-ray pool, which remains dominated by supernova remnants. As the shock wave expands and decelerates into the lower-density outer wind, the accelerated protons decouple from the shock and escape into the interstellar medium, contributing a localized, transient burst of cosmic-ray protons up to sub-TeV and few-TeV energies to the Galactic cosmic-ray sea, rather than acting as Galactic PeVatrons. Over secular timescales, recurrent eruptions carve a low-density cavity or nova super-remnant (NSR) around the binary; in \cite{LopezCotoNSR2025}, we showed that quiescent \textit{Fermi}-LAT upper limits towards this NSR constrain the local density to $n_{\rm ISM} \lesssim 0.1$\,cm$^{-3}$, confirming that novae inflate parsec-scale cosmic-ray bubbles without dominating Galactic cosmic-ray transport.
\end{itemize}

\section{Relativistic Jets in Microquasars: SS~433 and Beyond}
\label{sec:microquasars}

Microquasars are Galactic X-ray binary systems comprising a stellar-mass black hole or neutron star accreting mass from a stellar companion and launching collimated, relativistic bipolar jets. Operating on spatial and temporal scales millions of times smaller than supermassive black holes in active galactic nuclei, microquasars serve as compact factories of relativistic particles to investigate jet dynamics, particle acceleration, and non-thermal radiation mechanisms.

\begin{figure}[t]
    \centering
    \includegraphics[width=\columnwidth]{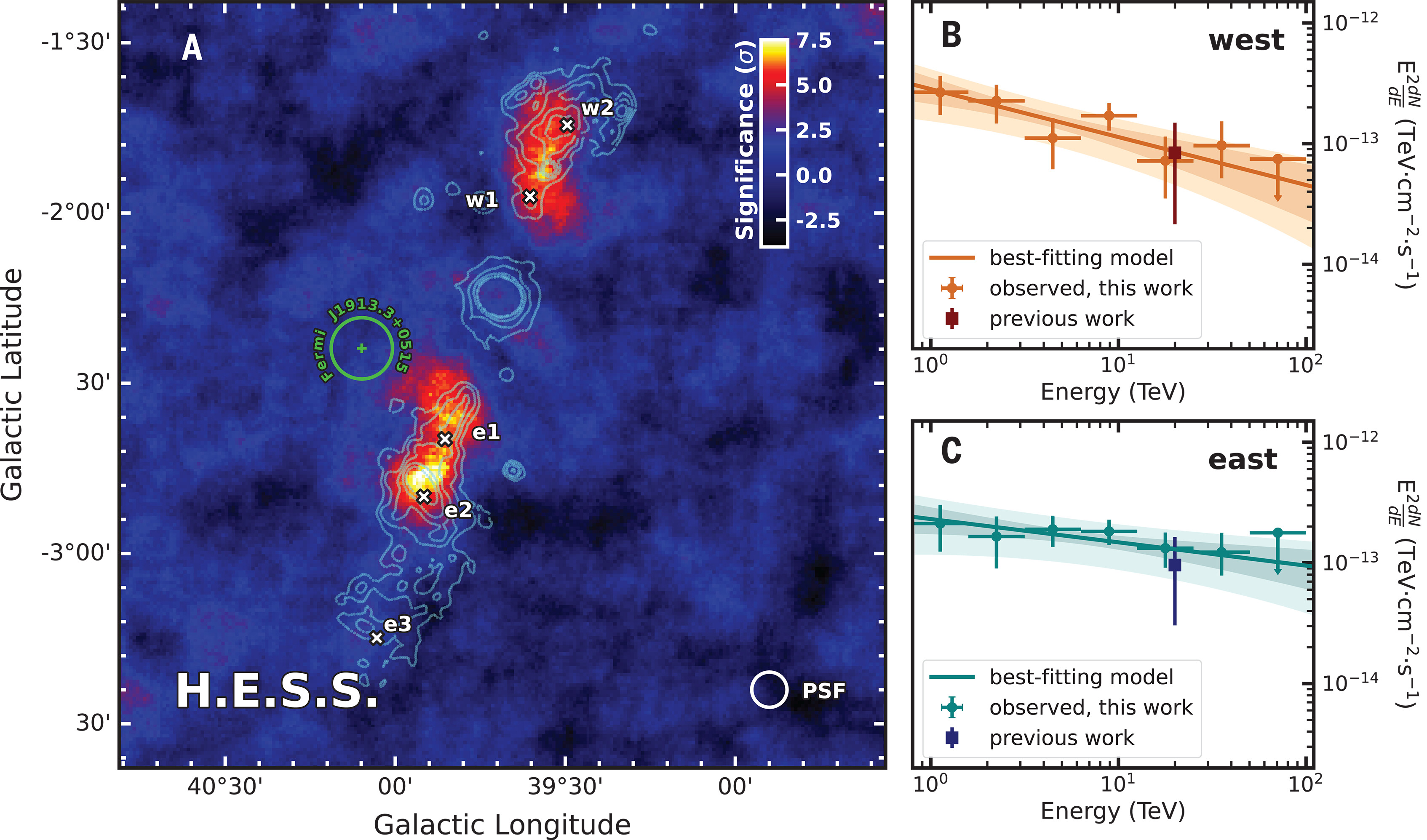}
    \caption{H.E.S.S. observations of the microquasar SS~433 / W50 system. Left: VHE $\gamma$-ray excess map above $1$\,TeV showing the spatial displacement of emission along the eastern (e1, e2) and western (w1, w2) jet lobes relative to the central binary (cross). Right: Measured energy spectra from the eastern and western interaction lobes, demonstrating multi-TeV emission extending up to tens of TeV. Original image from: \cite{HESSSS433_2024}.}
    \label{fig:ss433_hess}
\end{figure}

The archetypal microquasar SS~433 ($d \approx 5.5$\,kpc) features a black hole accreting matter at super-Eddington rates from an A-type supergiant donor, launching persistent, precessing baryonic jets ($v \approx 0.26c$, kinetic power $L_{\rm jet} \sim 10^{39}\text{ erg s}^{-1}$) that inflate and distort the surrounding W50 supernova remnant into the characteristic ``Manatee Nebula.'' Following the initial multi-TeV detection of the outer lobes by the wide-field HAWC observatory~\cite{HAWCSS433_2018}, H.E.S.S. dedicated over 200 hours of pointed observations to spatially and spectrally resolve the jet interaction regions in the $1$--$50$\,TeV band (Fig.~\ref{fig:ss433_hess})~\cite{HESSSS433_2024}.

The H.E.S.S. results revealed key physical properties of the system:
\begin{itemize}[leftmargin=*,itemsep=0ex,parsep=0ex,topsep=0.2ex]
    \item No VHE $\gamma$-ray emission is detected from the central binary position. This indicates that any VHE photons produced near the core are heavily absorbed via pair production ($\gamma\gamma \to e^+e^-$) in the intense radiation fields of the accretion disk and donor star, or that particle acceleration requires jet deceleration further downstream.
    \item Intense VHE emission emerges symmetrically from two distinct regions located $\approx 25$--$30$\,pc from the binary along both the eastern (lobes e1/e2) and western (lobes w1/w2) jets, precisely coinciding with non-thermal X-ray synchrotron regions resolved by \textit{XMM-Newton} and \textit{Chandra}. Both lobes exhibit hard multi-TeV spectra extending up to tens of TeV without an exponential cutoff.
    \item The emission exhibits an energy-dependent spatial gradient: the highest-energy photons ($E > 10$\,TeV) are concentrated near the shock fronts, whereas lower-energy photons ($0.8$--$2.5$\,TeV) extend further downstream. This morphology directly traces the radiative cooling (synchrotron and inverse Compton) of relativistic electrons as they advect away from the acceleration shocks. Broadband modeling confirms that electrons accelerated up to $\sim 100$--$200$\,TeV scatter Cosmic Microwave Background and infrared dust photons in magnetic fields of $B \approx 10$--$20\,\mu\text{G}$ under standard assumptions~\cite{HESSSS433_2024}.
\end{itemize}
In addition, LHAASO reported ultra-high-energy emission extending beyond $100$\,TeV from SS~433, with spatial correlation to surrounding molecular clouds, suggesting that an additional hadronic component may emerge at the highest energies where escaping hadrons collide with dense ambient gas~\cite{LHAASOSS433_2025}.

The success on SS~433 spurred targeted investigations of other Galactic microquasars. In V4641~Sagittarii ($d \approx 6.2$\,kpc, $\approx 6.4\,M_\odot$ black hole), H.E.S.S. resolved extended multi-TeV emission stretching $\approx 30$\,pc North and $\approx 55$\,pc South of the binary, bridging MeV--GeV emission and LHAASO measurements~\cite{HESSV4641_2026}. Conversely, extensive monitoring of the high-mass microquasar Cygnus~X-3 by MAGIC during major radio flares yielded no VHE detection~\cite{MAGICCygX3_2019}, placing stringent upper limits that constrain jet energetics, magnetic fields, and severe internal $\gamma\gamma$ opacity in compact Wolf-Rayet environments.

\section{Pulsars: The multi-TeV Challenge and Magnetospheric Physics}
\label{sec:pulsars}

Pulsars---rapidly rotating, highly magnetized neutron stars---convert rotational energy into relativistic plasma winds, acting as prime factories of relativistic electrons and positrons ($e^\pm$). For decades, classical pulsar electrodynamics relied on polar-cap or slot-gap models, where primary electrons emit curvature radiation along magnetic field lines within the corotating magnetosphere, predicting an unavoidable superexponential cutoff at $E \sim $a few\,GeV due to magnetic pair attenuation.

\begin{figure}[t]
    \centering
    \includegraphics[width=\columnwidth]{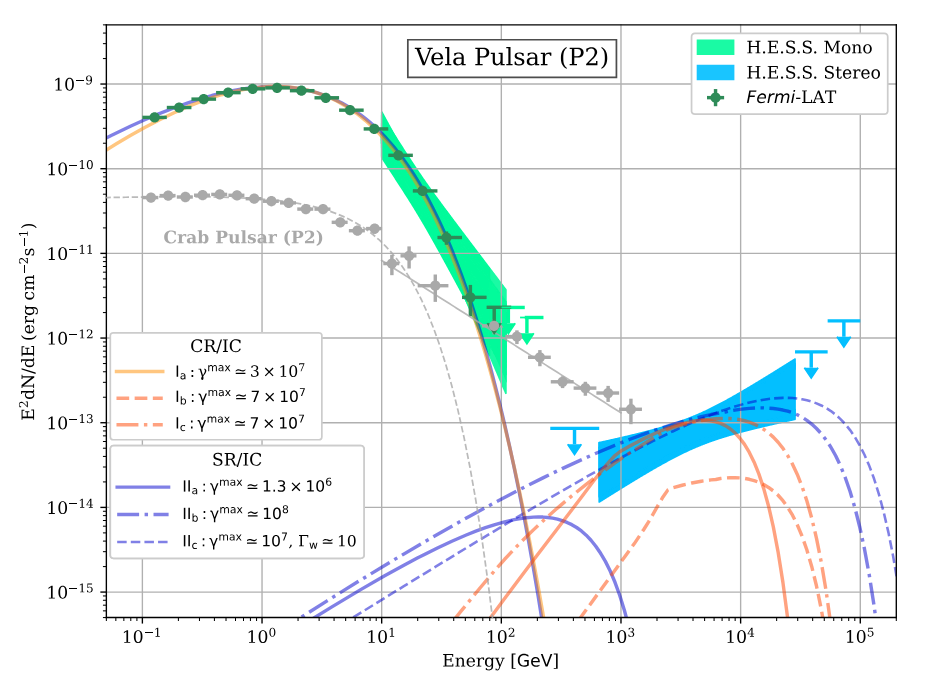}
    \caption{Broadband spectral energy distribution of the Vela pulsar's second pulse peak (P2), displaying the discovery by H.E.S.S. of a distinct, hard radiation component emerging above $100$\,GeV and extending up to $20$\,TeV. Green contours denote H.E.S.S. measurements; gray points show the Crab pulsar P2 spectrum for comparison. Original image from: \cite{HESSVela2023}.}
    \label{fig:vela_sed}
\end{figure}

This paradigm began to shift in 2008 when MAGIC detected pulsed emission from the Crab pulsar above $25$\,GeV~\cite{MAGICCrabPulsar2008}, followed by VERITAS and MAGIC measurements extending beyond $100$\,GeV and into the multi-hundred GeV regime~\cite{VERITASCrabPulsar2011}. Recently, LST-1 performed high-statistics observations of the Crab pulsar, resolving both the P1 and P2 peak profiles and confirming a power-law continuation up to $\sim 1$\,TeV~\cite{LSTCrab2024}. These observations established that VHE pulsed emission originates in the outer magnetosphere or current sheet, outside the light cylinder ($R_{\rm LC} = c/\Omega$), via inverse Compton scattering.

The frontier expanded further with observations of middle-aged pulsars. LST-1 detected pulsed VHE $\gamma$ rays above $15$\,GeV from the Geminga pulsar and PSR~B1706-44 ~\cite{LSTGemingaB1706_2025}. In Geminga, the spectrum shows a power-law tail without the sharp exponential cutoff characteristic of inner-magnetosphere curvature radiation, demonstrating that acceleration beyond the light cylinder is a generic feature across both young and middle-aged pulsars.

The most profound surprise in pulsar astrophysics was delivered by H.E.S.S. with the detection of the Vela pulsar (PSR~B0833-45)~\cite{HESSVela2023}. As shown in Fig.~\ref{fig:vela_sed}, Vela's primary pulse peak (P2) exhibits the expected sub-exponential cutoff at $\approx 3$\,GeV in \textit{Fermi}-LAT data. However, H.E.S.S. discovered an entirely separate, hard radiation component that emerges above $100$\,GeV and extends up to $20$\,TeV. This represents the highest-energy pulsed emission ever detected from any celestial body.

The detection of $20$\,TeV pulsed photons carries severe physical consequences:
\begin{itemize}[leftmargin=*,itemsep=0ex,parsep=0ex,topsep=0.2ex]
    \item At $20$\,TeV, $\gamma$ rays cannot escape the corotating magnetosphere without catastrophic pair production on background fields. The maximum photon energy attainable by curvature radiation in a dipole magnetosphere is limited by radiation reaction to $\sim 10$\,GeV.
    \item Producing $20$\,TeV photons via inverse Compton scattering requires electrons with Lorentz factors $\gamma_e \sim 10^7$. Such extreme acceleration cannot occur within the magnetosphere; it must take place in the unconfined relativistic striped pulsar wind, driven by magnetic reconnection in the equatorial current sheet well beyond the light cylinder ($R_{\rm LC} \approx 4.3 \times 10^7\text{ m}$ for Vela)~\cite{HESSVela2023}.
\end{itemize}
The recent detection by H.E.S.S. of a comparable multi-TeV pulsed component in PSR~J1509-5850 confirms that multi-TeV pulsed emission represents an emerging class of extreme particle accelerators, indicating that particle acceleration begins well before the pulsar wind reaches the termination shock of the surrounding nebula.

\section{The Galactic Center: PeVatron Status and Ridge Cutoff}
\label{sec:galactic_center}

The Galactic Center hosts the supermassive black hole Sagittarius~A* (Sgr~A*, $M \approx 4.1 \times 10^6\,M_\odot$), embedded within the Central Molecular Zone (CMZ), which contains dense gas masses of $\sim 3 \times 10^7\,M_\odot$.

\begin{figure}[t]
    \centering
    \includegraphics[width=\columnwidth]{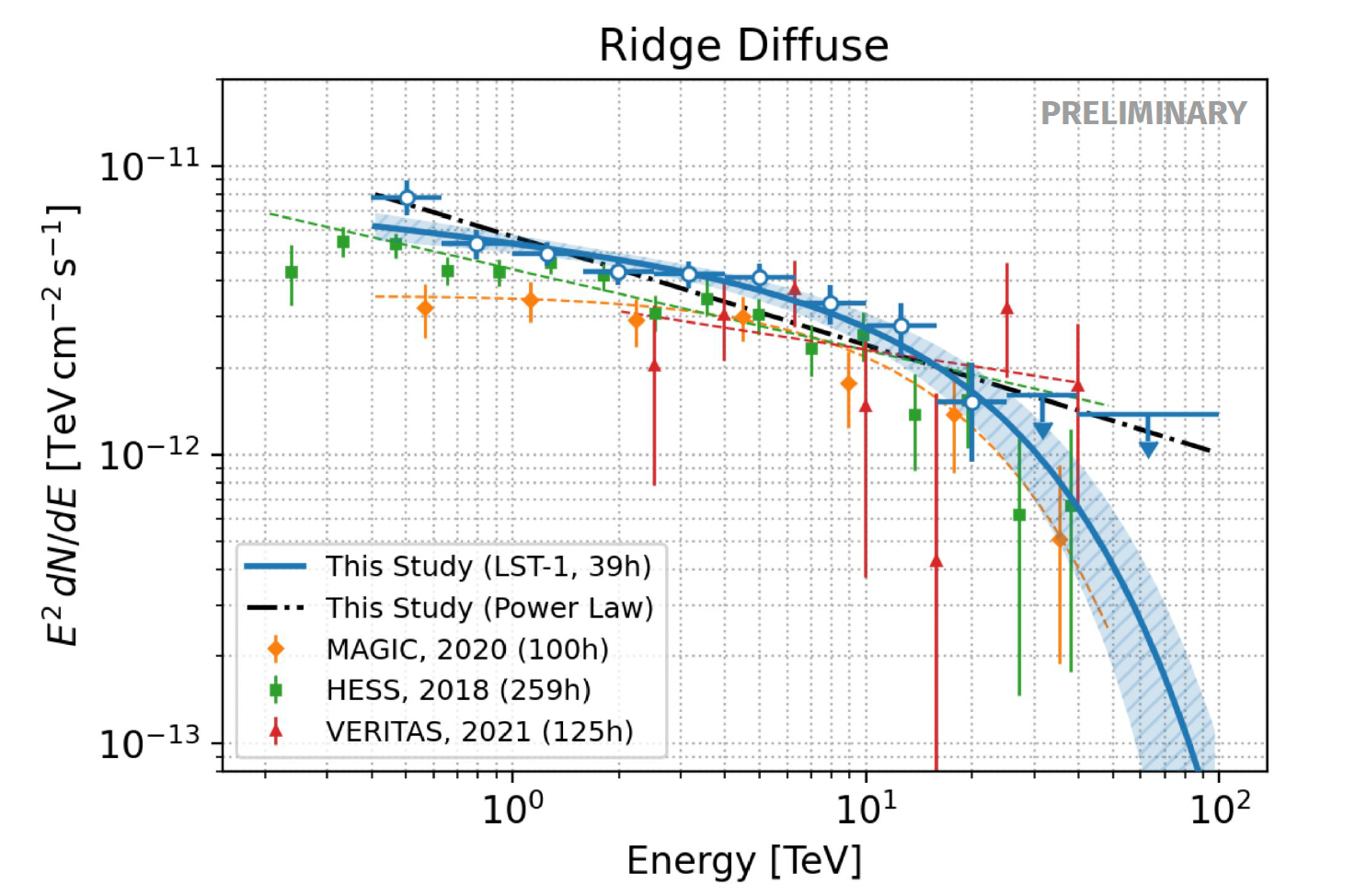}
    \caption{Spectral energy distribution of the Galactic Center ridge diffuse emission. Blue curves show the LST-1 measurements (39-hour exposure), displaying a preference for a spectral cutoff at $E_{\rm cut} \approx 24$\,TeV over a pure power law. Comparison datasets are shown from MAGIC, H.E.S.S., and VERITAS. Original image from: \cite{LSTGC2025}.}
    \label{fig:gc_ridge_sed}
\end{figure}

In 2016, H.E.S.S. discovered diffuse VHE $\gamma$-ray emission tracing the CMZ gas as an unbroken power law extending to $\approx 50$\,TeV without evidence of a cutoff~\cite{HESSPeVatron2016}. Since this emission arises from $pp$ collisions between cosmic rays and dense gas, the parent proton spectrum must extend up to at least $\approx 1$\,PeV, providing the first compelling evidence for a Galactic PeVatron. The cosmic-ray density profile ($w_{\rm CR}(r) \propto 1/r$) indicated quasi-continuous proton injection from a central source---likely linked to past activity of Sgr~A*---over the past $\gtrsim 10^3$--$10^4$\,years, diffusing outward with an energy-dependent diffusion coefficient $D(E) \propto E^\delta$.

To determine whether this spectrum continues into the PeV regime or terminates at tens of TeV, multiple IACTs performed deep follow-up campaigns. Fig.~\ref{fig:gc_ridge_sed} displays the composite CMZ diffuse spectrum, comparing 39 hours of LST-1 observations with archival data from H.E.S.S., MAGIC~\cite{MAGICGC2020}, and VERITAS~\cite{VERITASGC2021}. The high-precision LST-1 data indicate a preference for a spectral cutoff at $E_{\rm cut} \approx 24$\,TeV over an unbroken power law.

In parallel, a 15-year monitoring study of the central point source by H.E.S.S. found that its VHE flux is remarkably steady, with long-term variations constrained to $<28\%$~\cite{HESSGCLong2026}, ruling out major recurrent injection bursts over recent decades. The preference for a $24$\,TeV cutoff in the CMZ diffuse emission suggests that either the central accelerator operates with a maximum proton energy of $E_{p,\mathrm{max}} \sim 200$--$300$\,TeV, or that PeV protons accelerated in past active episodes have already diffused out of the inner CMZ ridge. The contrast between the steady VHE point source and dramatic flaring in X-ray and near-infrared bands further suggests that VHE emission originates in a larger, quasi-stationary circum-nuclear zone rather than the immediate accretion flow.

\section{Extragalactic Sky: Distant Blazars and the EBL Horizon}
\label{sec:extragalactic}

In the extragalactic sky, IACTs serve both as deep pointed instruments and as survey facilities. The H.E.S.S. Collaboration completed a systematic extragalactic survey covering $\approx 6\%$ of the Southern sky over $\approx 2700$\,hours, discovering 23 VHE sources and providing unbiased population constraints~\cite{HESSSurvey2025}, while a parallel survey by VERITAS added 5 new discoveries across $\approx 1000$\,hours.

\begin{figure}[t]
    \centering
    \includegraphics[width=\columnwidth]{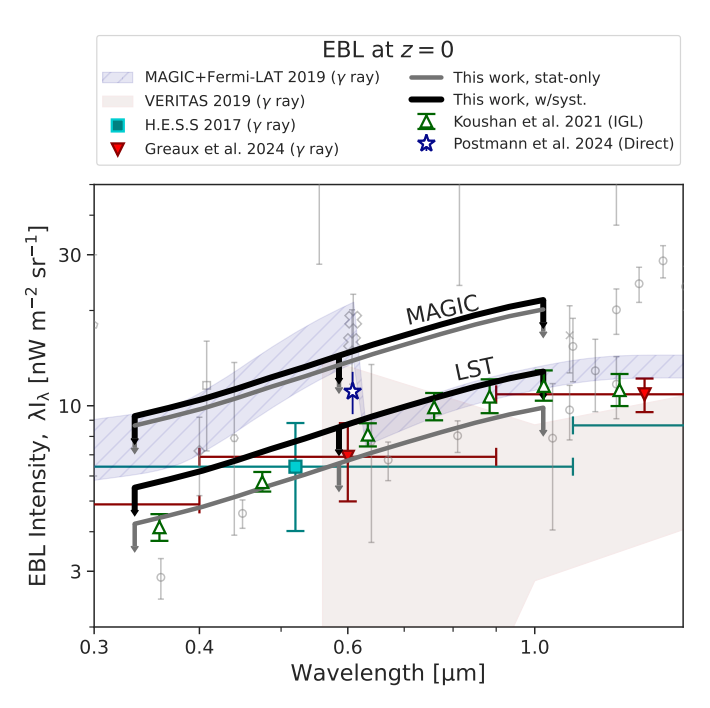}
    \caption{Constraints on the local Extragalactic Background Light (EBL) intensity at $z=0$ derived from LST-1 observations of the distant FSRQ OP~313 ($z = 0.997$), compared with constraints from MAGIC, VERITAS, and H.E.S.S., alongside integrated galaxy light lower limits and direct measurements. Original image from: \cite{LSTOP313_2026}.}
    \label{fig:op313_ebl}
\end{figure}

When VHE $\gamma$ rays travel across cosmological distances, they undergo pair production on optical-to-infrared photons of the Extragalactic Background Light (EBL), attenuating the intrinsic flux of the source.  VHE photons probe the optical and near-infrared EBL emitted by stars and dust throughout cosmic history. Above $100$\,GeV, the Universe becomes optically thick for sources at $z \gtrsim 0.5$, making the detection of distant flat-spectrum radio quasars (FSRQs) at $z \sim 1$ an exceptional challenge.

Two recent discoveries have broken this cosmological barrier:
\begin{itemize}[leftmargin=*,itemsep=0ex,parsep=0ex,topsep=0.2ex]
    \item \textbf{PKS~0346-27 ($z = 0.991$):} H.E.S.S. detected this distant FSRQ during a major optical/$\gamma$-ray flaring period~\cite{HESSPKS0346_2026}. Multi-wavelength monitoring revealed an unexpected absence of X-ray flaring and a $\approx 2$-day lag between the VHE peak and the \textit{Fermi}-LAT GeV peak. Standard single-zone leptonic models cannot reconcile this temporal offset and X-ray silence, favoring a hadronic synchrotron origin or an external Compton scenario in a structured jet.
    \item \textbf{OP~313 ($z = 0.997$):} In December 2023, following a \textit{Fermi}-LAT alert, LST-1 triggered rapid Target of Opportunity observations of OP~313, discovering VHE $\gamma$-ray emission at $>5\sigma$ significance~\cite{CortinaATel16381,LSTOP313_2026}. OP~313 is the most distant AGN ever detected by an IACT, extending the VHE observational horizon.
\end{itemize}
As illustrated in Fig.~\ref{fig:op313_ebl}, measuring the spectrum of OP~313 provides a direct cosmological probe of EBL opacity. Comparing the observed spectrum against EBL models~\cite{Dominguez2011} yields robust upper limits on the EBL photon density in the $0.3$--$1.0\,\mu\text{m}$ band, confirming that the EBL is dominated by resolved galaxy light and tightly constraining exotic contributions such as Population~III stars or axion-like particle conversions.

\section{Gamma-Ray Bursts: From Afterglow Physics to GRB~221009A}
\label{sec:grbs}

Gamma-ray bursts (GRBs) are the most energetic explosions in the Universe, produced by massive stellar core collapses (long GRBs) or compact object mergers (short GRBs). Over decades, ground-based searches yielded only upper limits due to slow telescope repositioning and high energy thresholds. This situation was revolutionized when MAGIC detected GRB~190114C ($z = 0.42$) above $1$\,TeV within 57 seconds of the alert~\cite{MAGICGRB190114C_2019}, followed by H.E.S.S. detections of GRB~180720B~\cite{HESSGRB180720B_2019} and GRB~190829A~\cite{HESSGRB190829A_2021}, and MAGIC's detection of GRB~201216C ($z = 1.1$). These observations established VHE emission as a standard component of GRB afterglows, driven by synchrotron self-Compton (SSC) radiation in external forward shocks expanding into the circumstellar medium.

\begin{figure}[t]
    \centering
    \includegraphics[width=\columnwidth]{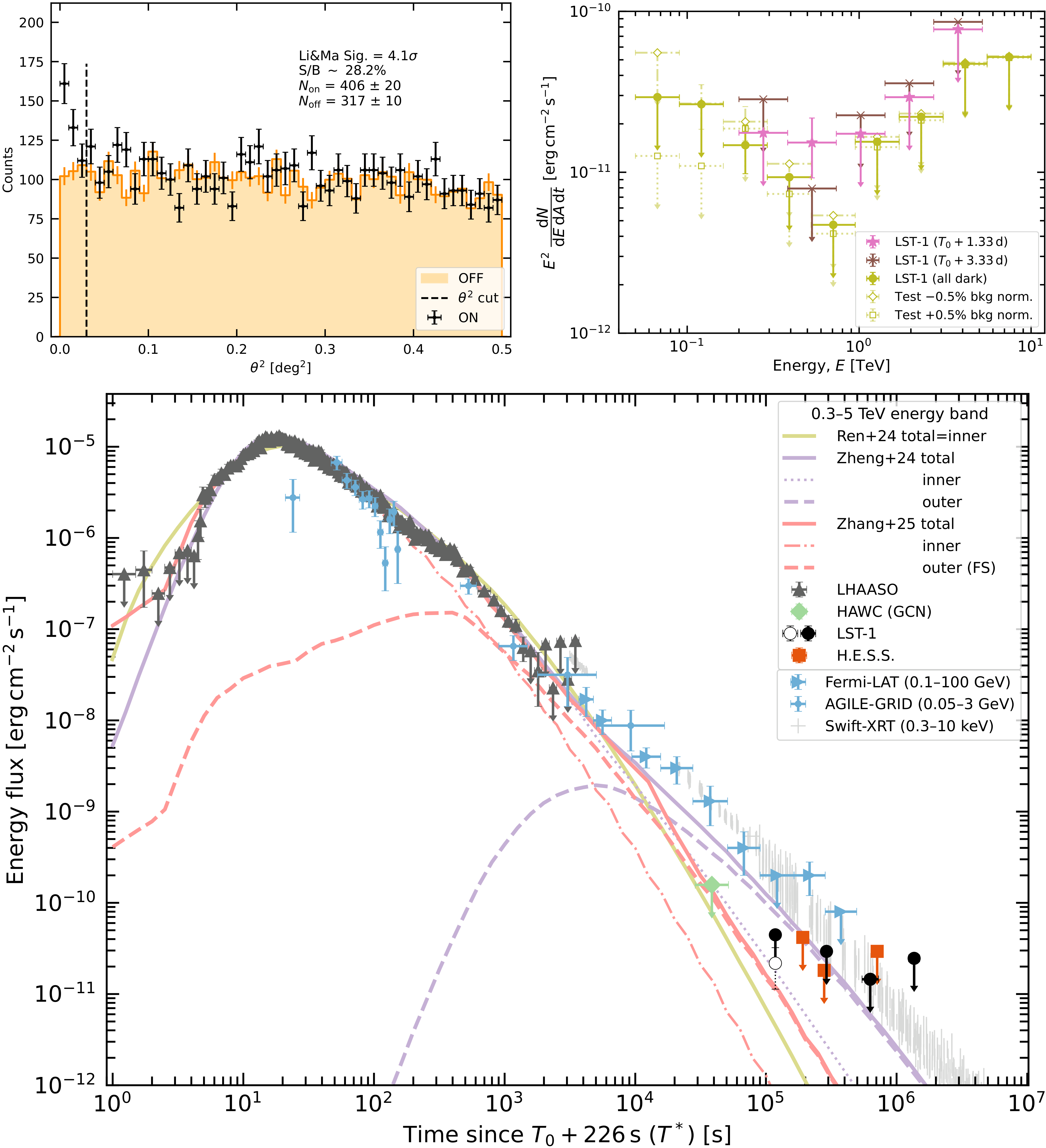}
    \caption{LST-1 observation of GRB~221009A (``The BOAT''). Top left: $\theta^2$ distribution showing the $4.1\sigma$ detection at $T_0 + 1.33$\,days. Top right: Differential energy flux points and upper limits. Bottom: Broadband afterglow light curve from radio to TeV confronting structured jet models. Original image from: \cite{LSTGRB221009A_2025}.}
    \label{fig:grb221009a}
\end{figure}

In classical shock acceleration theory, synchrotron emission cannot easily exceed the radiation-reaction burnoff limit. Detections above $100$\,GeV firmly demonstrate the emergence of a separate radiation component, predominantly inverse Compton scattering by electrons accelerated in the relativistic blast wave.

On 2022 October 9, the astronomical community witnessed GRB~221009A, the ``Brightest of All Time'' (the BOAT), with isotropic equivalent energy $E_{\rm iso} > 10^{55}$\,erg and peak flux exceeding $10^5$\,Crab units~\cite{LHAASOGRB221009A_2023}. At the time of the burst, observational conditions across IACT sites were hampered by strong moonlight. H.E.S.S. observed the field on October 11 ($T_0 + 2.2$\,days), deriving upper limits that constrained inverse-Compton cooling scenarios~\cite{HESSGRB221009A_2023}.

Remarkably, LST-1 initiated non-standard observations under intense moonlight on October 10 ($T_0 + 1.33$\,days), accumulating $\approx 4.1$\,hours of exposure. As shown in Fig.~\ref{fig:grb221009a}, LST-1 achieved a $4.1\sigma$ detection of the afterglow above $100$\,GeV~\cite{LSTGRB221009A_2025}. This measurement bridges the crucial temporal gap between early-time UHE detections by LHAASO ($t < 10^3$\,s) and late-time IACT upper limits ($t > 2$\,days), placing stringent constraints on structured jet models featuring an ultra-relativistic narrow core encased in a wider, slower cocoon.

\section{Fundamental Physics: Indirect Dark Matter Searches}
\label{sec:dark_matter}

IACTs provide powerful constraints on fundamental physics, notably in the indirect search for Weakly Interacting Massive Particles (WIMPs). If WIMP candidates ($\chi$) annihilate into standard model pairs ($b\bar{b}$, $\tau^+\tau^-$, $W^+W^-$), the resulting hadronization cascades generate a continuous spectrum of secondary $\gamma$ rays, alongside potential monochromatic lines ($\chi\chi \to \gamma\gamma$).

Dwarf spheroidal galaxies (dSphs) of the Milky Way are ideal dark matter targets due to their high mass-to-light ratios and lack of astrophysical $\gamma$-ray backgrounds. To maximize statistical power, the \textit{Fermi}-LAT, HAWC, H.E.S.S., MAGIC, and VERITAS collaborations conducted a joint likelihood analysis combining observations across all major $\gamma$-ray observatories~\cite{JointDMDsph2026}.

This unprecedented combined effort unites the complementary strengths of space-borne and ground-based arrays across WIMP masses from $5$\,GeV to $100$\,TeV. By combining dozens of dSphs simultaneously, the joint framework significantly reduces systematic uncertainties associated with individual $J$-factors. For WIMP masses below $\sim 100$\,GeV, the limits dive below the canonical thermal relic cross section ($\langle \sigma v \rangle_{\rm th} \approx 2.2 \times 10^{-26}\text{ cm}^3\text{ s}^{-1}$), robustly constraining standard thermal WIMP parameter space independently of astrophysical modeling of the Galactic Center.

Complementing dSph searches, H.E.S.S. published results from a 546-hour deep exposure of the Galactic Center halo searching for sharp monochromatic lines ($\chi\chi \to \gamma\gamma$)~\cite{HESSDMGC2026}. This study establishes the most stringent constraints on heavy WIMP line annihilation in the $500$\,GeV to $50$\,TeV mass range, reaching cross-section limits down to $\langle \sigma v \rangle \sim 10^{-28}\text{ cm}^3\text{ s}^{-1}$.

\section{Next-Generation Observatories and Methodological Advances}
\label{sec:outlook}

As current IACT facilities continue to produce landmark discoveries, the discipline is entering the era of the Cherenkov Telescope Array Observatory (CTAO). CTAO will deploy two full-sky arrays (CTAO-North on La Palma and CTAO-South in Chile), delivering an order of magnitude improvement in sensitivity, angular resolution, and spectral coverage ($20$\,GeV to $300$\,TeV)~\cite{CTAScience2019}.

Prototyping and early science validation are already well established. LST-1 is operating on La Palma, having produced the results on RS~Oph, Crab, Geminga, OP~313, and GRB~221009A highlighted here. Construction of LST-2, LST-3, and LST-4 at CTAO-North is advancing rapidly, heading toward four-telescope stereoscopic operations. Concurrently, novel Small-Sized Telescopes (SSTs), such as the ASTRI Mini-Array and the single-mirror SST-1M array (which recently demonstrated stereoscopic detection of the Crab Nebula at Ond\v{r}ejov~\cite{SST1M2025}), are paving the path for multi-TeV and PeV sky coverage.

Crucially, hardware advancements are accompanied by a revolution in reconstruction methodologies. Traditional Hillas parameterization is being superseded by deep learning frameworks such as CTLearn, which apply Convolutional Neural Networks (CNNs) and Graph Neural Networks (GNNs) directly to calibrated pixel charges and arrival times~\cite{CTLearn2021}. These models yield significant gains in background rejection and energy resolution, especially near threshold ($\lesssim 50$\,GeV) where shower images contain few Cherenkov photons, effectively lowering array thresholds and boosting survey sensitivity.

\section{Conclusions}
\label{sec:conclusions}

Over the past four decades, Imaging Atmospheric Cherenkov Telescopes have revolutionized our understanding of the high-energy Universe. The recent results summarized in this review demonstrate that ground-based $\gamma$-ray observatories continue to produce fundamental breakthroughs:
\begin{itemize}[leftmargin=*,itemsep=0ex,parsep=0ex,topsep=0.2ex]
    \item \textbf{Novae:} RS~Oph established recurrent novae as transient cosmic-ray proton accelerators up to sub-TeV energies in blast-wave wind shocks.
    \item \textbf{Microquasars:} Spatially resolved multi-TeV emission in SS~433 and V4641~Sgr lobes demonstrated electron acceleration in jet shocks and downstream advective cooling.
    \item \textbf{Pulsars:} Discovery of $20$\,TeV pulsed emission from the Vela pulsar demanded particle acceleration in the striped wind via magnetic reconnection.
    \item \textbf{Galactic Center:} Preference for a $24$\,TeV cutoff in the CMZ ridge clarified PeVatron proton diffusion and accelerator limits.
    \item \textbf{Extragalactic:} Detections of $z \approx 1$ blazars (OP~313, PKS~0346-27) provided tight cosmological constraints on the EBL.
    \item \textbf{GRBs:} Detection of GRB~221009A afterglow under moonlight by LST-1 probed structured jet physics.
    \item \textbf{Dark Matter:} Combined 5-instrument dSph and GC halo analyses achieved limits diving below the thermal relic cross section.
\end{itemize}
With the maturation of machine-learning reconstruction methods, synergies with wide-field arrays (HAWC, LHAASO, SWGO), and the deployment of CTAO, the future of ground-based VHE $\gamma$-ray astronomy promises an extraordinary era of discovery.

\clearpage
\begingroup
\renewcommand*{\bibfont}{\footnotesize}
\setlength{\bibitemsep}{0.05ex}
\printbibliography[title={References}]
\endgroup

\end{document}

%% file: gamma26-template.tex
\usepackage[british]{babel}
\usepackage{microtype}
\usepackage{amsmath,amsfonts,amsthm}
\usepackage[svgnames,table]{xcolor}
\definecolor{GammaDarkGreen}{rgb}{0.113,0.278,0.2745}
\definecolor{GammaCharcoal}{rgb}{0.0706, 0.0706, 0.0706}
\definecolor{GammaRed}{HTML}{B71325}
\definecolor{GammaGreen}{HTML}{004846}
\usepackage{booktabs}
\usepackage{lastpage}
\usepackage{graphicx}
\usepackage{enumitem}
\setlist{leftmargin=*,itemsep=0.4ex,parsep=0ex,topsep=0.8ex,partopsep=0ex}
\setlist[itemize]{label=\color{GammaRed}$\triangleright$}
\usepackage{sectsty}
\chapterfont{\normalfont\Large\color{GammaRed}}
\sectionfont{\normalfont\Large\color{GammaRed}}
\subsectionfont{\normalfont\large\color{GammaCharcoal}}
\subsubsectionfont{\normalfont\normalsize\color{GammaGreen}}

\usepackage{subfig}

\usepackage{caption}
\DeclareCaptionLabelFormat{abbrevfig}{\color{GammaRed}Fig.\ #2}
\DeclareCaptionLabelFormat{abbrevtab}{\color{GammaRed}Tab.\ #2}
\usepackage[pdftex]{hyperref}
\pdftrailerid{}
\hypersetup{colorlinks=true,linkcolor=GammaDarkGreen,citecolor=GammaDarkGreen,filecolor=GammaDarkGreen,urlcolor=GammaDarkGreen}

\addto\extrasenglish{%
}

\usepackage{geometry}
\usepackage[utf8]{inputenc}
\usepackage[T1]{fontenc}
\IfFileExists{nimbusmononarrow.sty}{\usepackage{nimbusmononarrow}}{}
\usepackage{tgheros}
\usepackage{sansmath}
\usepackage{fancyhdr}
\usepackage{titling}

\newcommand{\authorstyle}[1]{{\large\color{GammaCharcoal}#1}}
\newcommand{\institution}[1]{{\footnotesize\itshape\color{GammaDarkGreen}#1\/}}

\pretitle{%
    \noindent
    \begin{minipage}[t]{0.63\textwidth}\vspace{20pt}
        \LARGE\color{GammaRed}
}

\posttitle{%
    \end{minipage}%
    \hfill
    \begin{minipage}[t]{0.33\textwidth}\vspace{-50pt}
        \raggedleft
        \includegraphics[width=\linewidth]{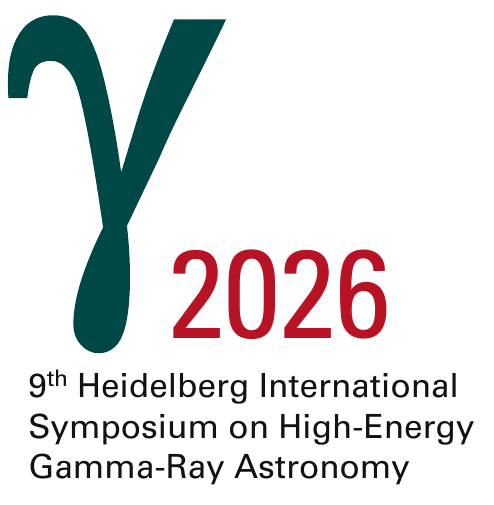}
    \end{minipage}%
}

\preauthor{\color{GammaDarkGreen}}
\postauthor{}
\predate{\begin{flushleft}\large\color{GammaRed}}
\postdate{\par\end{flushleft}}

\usepackage{lettrine}
\usepackage{xstring}
\newcommand{\initial}[1]{\lettrine[lines=2,findent=4pt,nindent=0pt]{\color{GammaRed}\emph{#1}}{}}
\newcommand{\lettrineabstract}[1]{\StrLeft{#1}{1}[\firstletter]\initial{\firstletter}\emph{{\StrGobbleLeft{#1}{1}}}}

\usepackage[backend=bibtex,style=numeric-comp,sorting=none,giveninits=true,citestyle=numeric]{biblatex}
\usepackage[autostyle=true]{csquotes}